# An application of cyberpsychology in Business Email Compromise (BEC)

Shadrack Awah Buo

Bournemouth University

Date: 19th June 2020

**Abstract**— This paper introduces Business Email Compromise (BEC) and why it is becoming a major issue to businesses worldwide. It also presents a case study of a BEC incident against Unatrac Holding Ltd and analyses the techniques used by the cybercriminals to defraud Unatrac Holding Ltd. A critical analysis of the psychological and sociotechnical impacts of BEC to both the company and employees are conducted, and potential risk mitigations strategies and recommendations are provided to prevent future attacks.

## 1. INTRODUCTION

Business Email Compromise (BEC) cases are increasing worldwide. The American International Group (AIG 2019) reported 23% of cyber claims were made by companies in Europe, the Middle East, and Africa. A report from the Federal Bureau of Investigation (FBI 2019) discovered that scammers stole $26 billion through BEC in 2016 and 2019. Further losses in the UK worth £32.2 million, according to the National Crime Agency (NCA 2018).

NCA (2018) defines BEC as a form of a phishing attack where the perpetrators impersonate senior employees. The aim is to deceive an employee to transfer funds or sensitive data without detection.

No business is immune to BEC attacks. This report analyses a case study of a UK affiliate of the US manufacturer, Caterpillar. Unatrac Holding Ltd lost over $11 million from the scheme, involving the offenders incorporating social engineering and providing fake invoices to fraudulently secure funds from the company's financial team. This report will analyze the techniques used, the psychological and sociotechnical impacts upon the business and employees and recommend solutions to prevent such attacks.

Additionally, this report argues that current technological solutions are inadequate in combating BEC, and significant emphasis should be placed on changing attitudes and behaviours. This report concludes with how BEC will develop in the future, due to technological advancements in Artificial Intelligence (AI).

## 2. INCIDENT IDENTIFICATION

In April 2018, the chief financial officer (CFO) of Unatrac received a phishing email. In an affidavit published by the Court listener (CL 2019), the email contained a link that led to a website page, crafted as a legitimate Microsoft Office 365 (MO365) login page. Believing the page was genuine, the CFO entered his login details which were captured by the intruders. They successfully gained access to the contents in the CFO's MO365 account, which included digital files such as tax filings and email templates.

The main culprit, Obinwanne Okeke, used the CFO's email account to issue fund transfers to the Unatrac's financial department. The scam involved sending fake invoices, which included Unatrac's logo and preformatted templates found in the compromised account to make the email appear legitimate (CL 2019). Armed with the knowledge that Unatrac invoices usually originate from outside the company's internal communications, the attackers modified the CFO's email settings to intercept communication between employees on the financial team. To conceal the intrusion from the victim, the attackers marked emails incoming as read and moved them to different folders (CL 2019). Unatrac's financial department issued 15 payments, totalling $11 million to multiple bank accounts controlled by the attackers.

## 2.1. MOTIVATIONS AND INTENTS

Cressey 's Fraud Triangle Theory (cited by Flores et al. 2018) provides factors that motivate cybercriminals to perform fraud. The factors include pressure, rationalization, and opportunity. This theory can be applied to this incident to understand the attacker's motivations.

A potential motivation was the perceived financial pressure on Okeke. In a BBC interview (2018), Okeke revealed that he grew up in poverty and his mother struggled to raise him and his siblings. Living in this environment motivated Okeke to promise himself to become rich to look after his mother and make her proud. Financial motivations can be revealed in the attackers' movements, as they tried to gain access and steal confidential data, which can be utilized for identity fraud (CL 2019, p.3) and potentially sold for profits. Further evidence of financial motivations can be inferred from the affidavit (CL 2019, p.7) where it described Okeke as a businessman and displayed a lavish lifestyle of travelling frequently on Instagram. Maintaining this lifestyle potentially pressures and motivates Okeke to commit BEC fraud.

Another potential motivation was the opportunity to exploit vulnerabilities in Unatrac's system. The described incident implies that the company had weak internal controls and security. For example, assumptions can be made from the affidavit (CL 2019) that anti-phishing software was not present in the company's IT systems. Vulnerabilities such as this would allow attackers to easily infiltrate and execute the phishing attack undetected.

Also, Okeke and his accomplices potentially found ways to rationalize this attack. The concept of rationalization refers to the attacker's belief to justify unethical and criminal behaviour. As Lister (2007) noted, determining rationalization is difficult if the attacker's background is unknown. With knowledge of Okeke's background, living in poverty can be linked to potential rationalizations. For example, one can suggest that Okeke's rationalization is based on the Robin-Hood narrative (Cukier et al. 2007), where he defrauds millions from large companies to financially provide for his mother and himself.

## 2.2 . Attacker's skill level

Okeke has an undergraduate degree in computer science and experience in web development (BBC 2018). This explains his ability to communicate in cryptic messages and use multiple nicknames to avoid detection. The affidavit also revealed Okeke and his accomplices were involved in other fraudulent schemes (CL 2019, p.9), which explains the high level of skills and professionalism demonstrated in the attack.

A phishing attack was implemented, a social engineering technique that tricks victims to provide sensitive information (Whitty and Young 2016). The affidavit (CL 2019, p.4) details how the attackers crafted a fraudulent website to convince the victim to provide sensitive information, such as login credentials. Moreover, the affidavit also reveals extensive communication and coordination between the attackers. With previous experience, Okeke directed his accomplices to create features and layouts for the phishing website to ensure it functions and appears genuine. The attackers also tested their web designs, by sharing copies of the source code, to check for errors and make corrections (CL 2019, p.6). The details reveal Okeke's organisational skills and further demonstrates his experience to successfully execute the BEC attack.

## 3. INCIDENT ANALYSIS

In the case study, the attackers used phishing and psychological techniques to elicit the CFO's email credentials and persuade the financial department to transfer funds. Even though the original phishing email was not revealed in the affidavit, the attackers likely employed a technical support scam shown in Figure 1.

According to one of Cialdini's (2001) principles of persuasion, society usually obeys authority even over their personal beliefs. Figure 1, the intruders impersonated as a known authority using a logo of Microsoft coupled with detailed and contextualized information regarding security updates, which compelled the CFO to click on the malicious link. The attackers also used Unatrac's logo and the authority of the CFO's hierarchical position to dupe the financial team to process fake invoices (CL, p.2), which included using the company's logo and preformatted invoice templates. The financial team would be unlikely to doubt the legitimacy of the invoice, especially from senior members. The principle of "liking" can also be applied as financial team members may question the purpose of the fund transfer requests without further administration, however, the professional communication in the form of invoices would contradict any doubts.

Additionally, it is likely the intruders applied "threat language" to the phishing email. Threat language exploits the victim's cognitive process to digest information and incites fear and worry, which affects the human's ability to process information and potentially decision making (Putri and Handoko 2018). In Figure 1, the sender wrote, "messages will be placed on hold until the above action is taken". This can be interpreted as a threat as confidential communications are held hostage until they follow the demand for clicking the link provided. The language could further impact the victim's decision making as they have less time to consider the legitimacy of the email.

The attackers and the victims displayed some of the "Big 5" personality traits (John and McCrae 1992) which led to the success of this attack. For example, Okeke and his accomplices strongly displayed "openness", the tendency to be creative which is evident from the preparation to the successful execution of the attack. For instance, in the affidavit, Okeke and his accomplices prepared for the attack by fabricating a convincing Microsoft login page. With evidence of extensive preparation and his financial motivation, it can be argued that Okeke also displayed high-level "conscientiousness", the tendency to be self-disciplined, self-organize and goal-oriented. However, he does display signs of narcissism, which ultimately led to his arrest. For instance, the FBI (CL 2019, p.7-8) found numerous posts and photos boasting his travels on social media. Consequently, posting photos on social media (CL 2019, p.7-8) allowed the FBI to track his movements and illegal activities through IP logs which contributed as evidence.

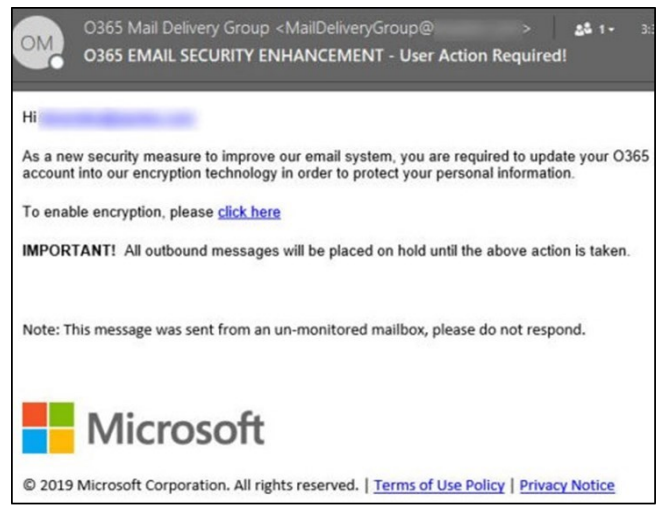

*Figure 1. Technical support scam (Nahorney 2019).*

### 3.1. Privacy and security concerns

The affidavit shows the devastating effects of the attacker's influence upon the victims, as they successfully convinced the CFO to click on the malicious link and capture his credentials which led to the defrauding of the company. Hadlington (2017) argues that adhering to cybersecurity procedures can be influenced by personality traits. For example, the CFO demonstrates a high level of extraversion personality traits in the form of impulsiveness due to his response to the phishing email. It is evident that impulsive individuals are more likely to conform to malicious requests like phishing attacks and are less risk-averse, such as failing to confirm the link's validity beforehand (Hadlington 2017).

Moreover, it is possible that cognitive overload could have impacted the decision making of the financial team. Kikerpill and Siibak (2019, p.55) noted that attackers use the principle of "scarcity" to create a sense of urgency. The desire for the financial team to urgently respond to a fraudulent email reduces their capacity to objectively evaluate the request (Kikerpill and Siibak 2019). Additionally, scammers deliberately send requests at the end of business hours when the staff are relaxed and less alert.

This attack resulted in a major breach of data privacy. With full access to the email account, the intruders could view sensitive information, such as tax filings and travel schedules. Also, copies of stolen passports and passwords of compromised accounts were found, which suggests that the attackers stole sensitive information to further break the law. Similarly, the attacker can sell stolen data for a profit which could lead to identity theft.

A potential security concern is the threat of ransomware attack. A common attack vector of ransomware is social engineering through phishing (Chen et al. 2017). The attackers could have infected Unatrac's network systems with malware and later demand a ransom, leading to further financial losses.

**4. FURTHER INCIDENT ANALYSIS**

Further analysis into the case study reveals the potential socio-technical and psychological impacts of BEC upon Unatrac and its employees. Modic and Anderson (2015) indicate that cybercrime can inflict a psychological impact on the victims. In this case study, the emotional impacts on the CFO can range from distress to emotions of betrayal, anger, and powerlessness (Watson et al. 2019). Also, victimization could lead to feelings of outrage, anxiety, and less trust in technology. In some instances, victims may develop a sense of shame and blame themselves for what happened (Nurse 2018).

Also, to the individual impacts of BEC, there are financial impacts for businesses. One of the impacts of BEC is the costs of repairs and security. The business needs to spend more money to restore the availability, integrity, and confidentiality of the compromised system and implement new security measures to prevent future attacks. These financial costs can potentially result in staff redundancies, which has a damaging impact on livelihoods. Also, BEC could damage the business reputation and customer attrition (Eeten et al. 2017, p.40), especially when the incidents are publicized. Damage to reputation can lead to the loss of market shares and expenses to repair public relations (Eeten et al. 2017, p.40).

**5. RISK MITIGATION AND RECOMMENDATIONS**

Organisations rely heavily on technology and awareness training campaigns to mitigate security risks (Eminağaoğlu et al. 2009), however, this cannot be an exclusive solution as they are not optimal in changing employee behaviour. Therefore, risk mitigation should shift focus towards changing the behaviour patterns of employees, rather than programs that do not guarantee full compliance. However, changing their conduct is more than educating them about reactive and risk behaviours, employees need to comprehend it and be motivated to apply the information in their everyday practice (Bada et al 2019).

This incident report can identify behavioural patterns in employees: impulsiveness, the lack of security protocols and persuaded by correspondence from those they perceive as the authority. Bada et al. (2019) argue that it is common for people to have security awareness but fail to apply what they learn in real life. For example, the potential lack of security protocols indicates that the company has not implemented security systems that could reduce the risk of phishing. Security and privacy practices are sometimes difficult to use, resulting in human error (Bada et al. 2019).

In terms of solutions, Bada et al (2019) argue that there are two methods to change behaviour: changing what people consciously think or shaping behaviour from automatic processes of judgement and influence without changing their thinking. The EAST framework reflects the second method of changing behaviour, which aims to deliver behaviour change by making targeted actions: "Easy, Attractive, Social and Timely" (Service et al. 2014). This framework is based on the Nudge theory, which highlights the use of positive reinforcements to influence behaviours (Halpern 2015). The framework can be applied to address the employee's behaviour by increasing alertness and reporting fraudulent emails.

The "Easy" aspect of the framework can be applied by providing frequent reminders about BEC attacks as well as providing examples of fraudulent emails, so the employees know what to look for (Service et al 2014). For instance, when the CFO log-in to his email account there should be a reminder to stay alert about phishing email. The aim of this step is not to transform employees into experts, but to help them stay alert and avoid phishing emails. Additionally, the "Attractive" aspect can be used to make security reminders engaging to the user. For example, the reminders should be tailored to the employees. It should explain the potential impacts to the company and the user by using a real-world example of similar attacks and their knock-on effects.

The "Timely" aspect of the framework can be applied to prevent impulsive decisions. This aspect prompts users when they are likely receptive, allows them time to consider the potential consequences and plan their responses (Service et al 2014). Contextually, before the CFO clicked the malicious link, a visual prompt could appear at the "right moment" informing the user of the risk indicators, such as the "sender is unknown". This allows the user to reflect on his decision and proceed to determine the legitimacy of the email.

Moreover, the aspect of "Social" can be used to encourage employees to flag or report malicious emails. According to Cialdini (2001), individuals are easily influenced by their peers. Therefore, this can be utilised to highlight suspicious emails reported by other employees and in return, employees should be given feedback for maintaining a secure behaviour (Service et al 2014).

**5.1 . Technical recommendations**

The most crucial step in mitigating phishing attacks is early detection. Multiple tools have been developed to automate the detection process. Meyers (2018) suggests implementing anti-phishing tools to prevent attacks. Tools, such as SpoofGuard can be used to alert users of malicious sites. However, studies by Zeydan et al. (2014) reveals that anti-phishing tools are not completely reliable, therefore, should not be used as an exclusive solution. The use of Domain-based Message Authentication Reporting and Conformance (DMARC) can prevent spoofed emails domains, although this will not protect against fraudsters with legitimate email domains (Mansfield-Devine 2016).

Blacklists and whitelists are other solutions that may protect BEC. Blacklists allow the user to regulate communication from unwanted senders (Siadati et al. 2016). For example, when the attacker's IP or email address is added to a blacklist, the connection is eliminated and prevents the email being received. On the other hand, whitelists consist of trustworthy email contacts. Although blacklists could help prevent attacks (Steer 2017), there should be a balance to ensure they intercept only fraudulent emails, and not restrict genuine communications which could affect the business.

**5.2. Non-technical recommendations**

Binks (2019) argues that training and awareness is an effective countermeasure for BEC. Agarwal (2019) recommends phishing simulation to display examples of phishing attacks and the impact it can have on the business. These training programs can help the employees recognise the techniques and cues deployed in BEC attacks (Ross, 2018). However, a drawback of awareness and training is the conflict between employee behaviours and organisational goals with security policies. The business goal of making profits can cause employees to emphasise performance rather than compliance, for example, an employee may take decision-making shortcuts to reach goals

(Zweighaft 2017, p.5).Therefore, employee awareness and training may become ineffective within the organisation where risk-mitigation fails to align with their goals (Fineberg 2014).

Company policies and processes can play a crucial role in preventing BEC. Governance processes should be established to help employees verify and validate all payments requests consistently and securely (Mansfield-Devine 2016). For example, phone calls to the financial team can verify the validity of payment requests. Also, policies that require multiple employees to process large transactions (Meyers 2018). Additionally, employees should observe financial accounts for anomalies and fraudulent activities.

**6. FUTURE CHALLENGES**

BEC symbolizes a continuous challenge with businesses as methods are becoming increasingly sophisticated, inflicting crippling financial losses. Technological solutions have been recommended to prevent further attacks; however, these should not be relied upon to prevent BEC. Similarly, training and awareness is not an exclusive solution either. Rather, a combination of human and technical solutions should be implemented to prevent attacks.

In terms of future challenges, Symantec (2019) predicts that attackers will soon deploy AI due to the evolution of machine learning. They argue that AI can be utilized to produce "deepfakes", a combination of visual and audio depictions of CEOs and other senior members of companies. Incorporating both visual and audio cues in their approach can enhance the fraudster's claim of legitimacy. Currently, there are no solutions to prevent this attack vector. Therefore, research is crucial to explore both human and technological factors to effectively combat BEC fraud in the future.